\documentclass[12pt,preprint]{aastex}

\def\be{\begin{equation}}
\def\ee{\end{equation}}
\def\eng{{\cal E}}
\def\vel{\vartheta}
\def\ga{\gamma}
\def\la{\lambda}
\def\md{\dot{\cal M}}
\def\sig{\Sigma}
\def\lsim{\lower.5ex\hbox{$\; \buildrel < \over \sim \;$}}
\def\gsim{\lower.5ex\hbox{$\; \buildrel > \over \sim \;$}}

\shorttitle{Standing Shocks Around Black Holes}
\shortauthors{S. Das, I. Chattopadhyay and S.K. Chakrabarti}

\begin{document}


\title{Standing Shocks around Black Holes: An Analytical Study}


\author{Santabrata Das, Indranil Chattopadhyay and Sandip K. Chakrabarti\altaffilmark{1}}
\affil{S. N. Bose National Centre for Basic Sciences, JD-Block, Salt Lake,
Kolkata, 700098, INDIA\\
e-mail:  sbdas@boson.bose.res.in, indra@boson.bose.res.in \& chakraba@boson.bose.res.in}


\altaffiltext{1}{Honorary Scientist, Centre for Space Physics,
114/v/1A Raja S.C. Mullick Rd., Kolkata 700047, INDIA}

\begin{abstract}
We compute locations of sonic points and standing shock waves
in a thin, axisymmetric, adiabatic flow around a Schwarzschild black hole.
We use completely analytical method to achieve our goal. Our results
are compared with those obtained numerically and a good agreement is seen.
Our results positively prove the existence of shocks in centrifugal pressure
dominated flows. We indicate how our results could be used to
obtain spectral properties and frequencies of shock oscillations
which may be directly related to the quasi-periodic oscillations of hard X-rays.
\end{abstract}

\noindent {\bf Astrophysical Journal; In press (August 20th issue)}


\section{Introduction}

In recent years, study of standing and oscillating shocks in accretion
flows has become very important since it is recognized that the spectral states of
black holes as well as Quasi-Periodic Oscillations (QPOs) observed in
light curves of black hole candidates are directly related to the radiative
transfer properties of a compact Comptonizing region close to a black hole
(e.g., Chakrabarti and Titarchuk 1995, CT95; Ling et al. 1997; Chakrabarti \& Manickam, 
2000, CM00; Muno et al., 2000; Feroci et al. 2000; Homan et al., 2001).
These shocks have been studied most extensively in the
literature (Chakrabarti 1989, C89; Chakrabarti 1990, C90)
and their properties have been verified by several independent groups 
\citep{yk95,nh94,lff97} of workers. If for a given set of initial parameters 
the standing shocks exists analytically, numerical simulations also would
find them \citep{cm93,mlc94,r95,mrc96} otherwise, the shock 
would be oscillating \citep{msc96,rcm97} causing
QPOs in the X-rays emitted from the post-shock region. Furthermore,
it has been observed that at least in some of the black hole
candidates, such as GRS1915+105, outflows are 
produced from the region which also emits the Comptonized
photons \citep{f00,d00} i.e., the post-shock
region or the centrifugal barrier dominated region according to
our present understanding of the flow solutions.

It is therefore pertinent to ask if the properties of the shocks,
such as location, strength, compression ratio etc. could be
understood solely analytically. The solutions obtained so far by
Chakrabarti and his collaborators and other groups always
resort to numerical means. Roughly, the method was the following:
for a given set of parameters (such as the specific energy and specific angular
momentum), it is first determined if the flow allows more than one
X-type sonic points (C89). This is because, at the horizon, matter must
have velocity same as that of the velocity of light and for
causality arguments, the flow must be supersonic. Thus, once the
flow becomes supersonic through the outermost sonic point,
and forms a shock (i.e., jumps to a subsonic branch) it must
pass through the inner sonic point to become supersonic at the
horizon. Next question is whether the specific entropy at the
inner sonic point is higher compared to that at the outer sonic
point. This is because, at the shock, entropy must be generated
and post-shock flow which passes through the inner sonic must
have higher entropy. The final and the most important
question is whether all the Rankine-Hugoniot conditions
(also known as the shock conditions) are satisfied somewhere
between the two X-type sonic points. The three Rankine-Hugoniot
conditions could be combined to obtain a combination of
Mach numbers (Shock Invariant). C89 obtained this expression
which is continuous across the shock and used this to obtain
the shock location by iterative technique.

In the present paper, we follow the same philosophy,
but obtain the shock locations analytically. There were two
motivations to do this.  First, from the theoretical point
of view, it is challenging to find solutions of a large
number of non-linear equations which must satisfy
a number of conditions mentioned above. Second, from the
observers' point of view, any observation which could
require standing shock waves, could be explained using
more fundamental parameters, such as specific energy
and angular momentum or even better, using accretion
rates of Keplerian and sub-Keplerian flows as in a
two-component flow solution of CT95. 
Analytical work also gives insight into why the
shocks form in the first place. The boundary of the
parameter space for shock formation is also obtained by
analytical means. These findings are important as
they would tell us when QPOs may or may not be seen.
These details would be discussed in future.

In the next Section, we present the model equations and
shock conditions. In \S 3, we present the sonic point
analysis and write down the expression for Shock
Invariant (C89) for the shake of completeness. In \S 4,
we present the analytical expressions of the sonic
points and discuss how the parameter space is divided
into regions of one or three sonic points.
In \S 5, we present expression of shock locations and again divide 
the parameter space into regions which may or may not have shocks.
We compare our solutions with numerical work.
In \S 6, we discuss some of the astrophysical implications
of our solutions and finally, in \S 7, we make concluding remarks.

\section{Model Equations and Shock Conditions}

We start with a thin, axisymmetric, inviscid, steady flow which is in equilibrium
in a direction transverse to the flow.
The model equations which govern the motion of matter accreting into
black hole are given as follows (C89):

\noindent {\it A. Radial Momentum Equation:}
\be
\vel \frac{d\vel}{dx} + \frac{1}{\rho} \frac{dP}{dx} + \frac{\la^2}{x^3}
+\frac{1}{2(x-1)^2} = 0
\ee
In a non-viscous flow, integration of this equation leads to the
{\it Energy Conservation Equation :}
\be
{\cal E} = {\frac{\vel_{e}^{2}}{2}}+{\frac {a_{e}^{2}}{\ga - 1}}
+{\frac {\la^{2}}{2x^{2}}}+g(x)
\ee
where, $g(x)$ is the pseudo-Newtonian potential introduced
by Paczy\'nski \& Wiita (1980) which is given by,
$g(x) = -{\frac {1}{2(x-1)}}$. Here, $\rho$ is the mass density, $P$
is the isotropic pressure, $\vel_{e}$ and
$a_{e}$ are the non-dimensional radial and the sound
velocities measured in units of velocity of light $c$,
$x$ is the non-dimensional radial distance measured in
units of the Schwarzschild radius $r_{g}={2GM/c^{2}}$,
$M$ being the mass of the black hole and $G$ being the
gravitational constant. $\gamma$ is the adiabatic index of the
flow and $P=K\rho^\gamma$ is assumed to the equation of state. $K$ is the
measure of the specific entropy which is constant except at the
shock location where local turbulence generate some entropy enabling the flow
to pass through the inner sonic point. The subscript $e$ refers to
the quantities measured on the equatorial plane. \\

\noindent {\it B. Continuity Equation:}
\be
\frac{d}{dx} (\vel_{e} \rho x h) = 0
\ee
which is integrated to obtain the {\it Mass Conservation Equation :}
\be
\dot{M} = \vel_{e} \rho  x h,
\ee
where, $h$ is the half-thickness of the flow at radial
coordinate $x$. Here ${\dot {M}}$ is the mass accretion rate
apart from a geometric constant. If we assume that the flow is in hydrostatic
equilibrium in the transverse direction, the
vertical component of gravitational force balances
the pressure gradient force. Hence the expression
for the half thickness of the disc is given by (C89),
\be
h(x)= a_{e}x^{1/2}(x-1).
\ee
We write the mass flux conservation equation in terms of $\vel_{e}$
and the sound speed on the equatorial plane $a_{e}=\sqrt{\ga P/\rho}$
in the following way,
$$
\md = \vel_{e} a_{e}^{q}x^{3/2}(x-1) =\vel_{e} a_{e}^{q} f(x)
\eqno{(4a)}
$$
\noindent where, $q =  {\frac {\ga +1}{\ga- 1}}$ and $f(x)=x^{3/2}(x-1)$.

The shock conditions which we employ here are the following (C89): \\

\noindent (a) the energy flux is continuous across the shock ---
\be
\cal E_{+} = E_{-},
\ee

\noindent (b) the mass flux is continuous across the shock ---
\be
{\dot{M}}_{+} ={\dot {M}}_{-}
\ee

\noindent and finally, (c) the momentum balance condition ---
\be
W_{+}+\sig_{+} {\vel^{2}_{e}}_{+} = W_{-}+\sig_{-} {\vel^{2}_{e}}_{-}
\ee
where subscripts ``$-$'' and ``$+$'' refer, respectively,
to quantities before and after the shock.
A shock satisfying these conditions is termed as a
Rankine-Hugoniot shock \citep{ll59}. Here, $W$ and
$\sig$ denote the pressure and the density, integrated
in the vertical direction [see, e.g., \citet{mk84}], $i.e.$,
\be
\sig = \int \limits_{-h}^{h}\rho dz = 2\rho_{e}I_{n}h,
\ee
and
\be
W = \int \limits_{-h}^{h}P dz = 2P_{e}I_{n+1}h,
\ee
where, $I_{n} = \frac  {(2^{n}n!)^{2}}{(2n+1)!}$,
$n$ being the polytropic index as defined previously.
In the subsequent analysis we drop the subscript $e$
if no confusion arises in doing so.

\section{Sonic Point Analysis and Shock Invariants}
In order to have a shock, the flow must be supersonic
i.e., the stationary flow must pass through the sub-sonic
flow to a supersonic flow. Discussions in this Section are
based on early works [see, \citet{c89,c90}].

\subsection{Sonic Point Conditions}

From the radial momentum equation and the continuity equation (Eqs. 1 and 3)
we derive the sonic point condition (or critical point condition)
in the usual way (C89). The first derivative of the radial velocity w.r.t
radial distance is given by,
\be
\frac {d\vel}{dx} =
\frac {\frac {2a^{2}}{\ga+1}\frac {d ln(f)}{dx}-\frac {dG}{dx}}
{{\left[ \vel -\frac {2a^{2}}{(\ga +1)\vel}\right]}}.
\ee
Here, $G(x)= \frac {\la ^{2}}{2x^{2}}-\frac {1}{2(x-1)}$ is the effective
potential. Since the denominator must vanish at the sonic points, if the
flow is assumed to be smooth everywhere, the numerator  must also
vanish simultaneously. The vanishing of the denominator gives,
\be
\vel^{2}_{c}(x_{c})=\frac {2}{(\ga +1)}a^{2}_{c}(x_{c}).
\ee
The factor (which is unity only in isothermal flows) in front of $a^{2}_{c}(x_{c})$ 
arises because the flow is assumed to be in vertical
equilibrium. The vanishing of the numerator gives,
\be
a^2_c(x_c)=\frac {(\ga+1)(x_c-1)}{x^2_c} \frac {[\la^2_K(x_c)-\la^2]}{(5x_c-3)}.
\ee
The subscript $c$ denotes quantities at the critical points.
Here $\la_K$ is the Keplerian angular momentum defined as
$\la^2_K=x^3_c/{[2(x_c-1)^2]}$. It is to be noted that since
the left hand side is always positive, angular momentum at the sonic
point must be sub-Keplerian, i.e., $\lambda (x_c) <\lambda_{\it K}$
(e.g., C90 and references therein).

\subsection{Mach Number Relation at the Shock}

From the equations given in \S 2, we now seek an invariant relation which must
be satisfied at the shock (C89). We rewrite the condition of
energy flux continuity (Eq. 6), and the pressure balance condition (Eq. 8)
in terms of the Mach number $M=\vel/a$ of the flow,
\be
\frac {1}{2}M^{2}_{+}a^{2}_{+}+\frac {a^{2}_{+}}{\ga -1}=
\frac {1}{2}M^{2}_{-}a^{2}_{-}+\frac {a^{2}_{-}}{\ga -1} ,
\ee
\be
{\dot {\cal M_{+}}}=M_{+}a^{\nu^{'}}_{+}f(x_{s}) ,
\ee
\be
{\dot {\cal M_{-}}}=M_{-}a^{\nu^{'}}_{-}f(x_{s}) ,
\ee
where
$\nu^{'}=\frac {2\ga}{\ga-1}$, and 
\be
\frac {a^{\nu}_{+}}{\dot {\cal M}_{+}}{\left( \frac{2}{3\ga-1}+
M^{2}_{+}\right)}=\frac {a^{\nu}_{-}}{\dot
{\cal M}_{-}}{\left( \frac{2}{3\ga-1}+M^{2}_{-}\right)} ,
\ee
where, $\nu = \frac {3\ga -1}{\ga -1}$ and $x_{s}$ is the location
of the shock. From Eqs. (14-17) one obtains the following equation
relating the pre- and post-shock Mach numbers of the flow at the shock (C89),
\be
C=\frac {\left[ M_{+}(3\ga-1)+(2/M_{+})\right]^{2}}{2+
(\ga -1)M^{2}_{+}}=\frac {\left[ M_{-}(3\ga-1)+(2/M_
{-})\right]^{2}}{2+(\ga -1)M^{2}_{-}} .
\ee
The constant $C$ is invariant across the shock. The Mach number of the flow just
before and after the shock can be written down in terms of $C$ as,
\be
M^{2}_{\mp}=\frac {2(3\ga-1)-C \pm \sqrt {C^{2}-8C\ga}}{(\ga-1)C-(3\ga-1)^{2}} .
\ee
The product of the Mach number is given by,
\be
M_{+}M_{-}=-\frac {2}{\left[ (3\ga-1)^{2}-(\ga-1)C\right]^{1/2}}.
\ee

\section{Analytical Expression of the Sonic Points and Behavior in Parameter Space}

To obtain shock locations, we need to first obtain the locations of the
sonic points, and ensure that at least two of them are X-type (C90).
In \S 3.1, we presented the sonic point conditions.
Using the definition $M=\vel/a$ of the Mach number,
and substituting $a^2_c(x_c)$ from Eq. (13) into Eq. (2),
we obtain following algebraic equation for $x_c$ given by,
\be
{\cal N}x^{4}_{c}-{\cal O}x^{3}_{c}+{\cal P}x^{2}_{c} \nonumber \\
-{\cal Q}x_{c}+{\cal R}=0.
\ee
\noindent where,
${\cal N} = {10\eng}$;
${\cal O}= {16\eng+2n-3}$;
${\cal P}={6\eng+\la^2(4n-1)-3}$;
${\cal Q}={8n\la^2}$;
${\cal R}={(1+4n)\la^2}$
and $n$ $(=\frac {1}{\ga -1})$ is the polytropic index.

We solve this equation analytically \citep{as70}
and get the location of the sonic points. Details are given in Appendix A.
For the purpose of critical or sonic points, $D$ of the
Appendix would be denoted as $D_c$. The equation has
four roots and Eq. A.8 can be used to check whether all
of them are complex (pair of complex conjugates) or at least two are real
and two are complex or all four are real. {\it A necessary condition to
form a shock wave is to have all four real roots.} Of course, only one would
be inside the black hole, and the other three would be outside, and out of these,
due to topological reasons, only two would be X-type or saddle type and
the one in between must be `O'-type or Center type. These
are determined by computing derivative $d\vel/dx$ at the sonic point by using
L'Hospital's rule and checking if they are real. For our purpose, two derivatives
at each sonic point must be real and of opposite signs in order that the
sonic point be of X-type or saddle type.

Figure 1 shows the division of the parameter space.
Denoting the discriminant $D$ (of Appendix A) by
$D_c$ we find that the condition $D_c<0$ is the
necessary condition to have three sonic points.
The boundary $D_c=0$ separates this region on which
two sonic points merge and the third one remains
separate. Outside of this region $D_c>0$, and only
one sonic point is possible and the other two roots
are complex conjugate of each other. The dotted
curve in the middle represents the condition:
$$
{\dot {\cal M}}_o = {\dot {\cal M}}_i ,
$$
where ${\dot {\cal M}}_i$ and ${\dot {\cal M}}_o$ are the entropy 
accretion rates at the inner and outer
sonic points respectively.  The region above it contains parameters
with  ${\dot {\cal M}}_i < {\dot {\cal M}}_o$ and
the region below it contains parameters
with ${\dot {\cal M}}_i > {\dot {\cal M}}_o$. This
latter region is suitable for shock formation in
accretion flows.

In passing we wish to point out that from Eq. (11) one can easily show
that locations where $dM/dx=0$ exactly coincide with the sonic points of the
flow. Thus the number of extrema of $M=M(x)$ is the same as that of the sonic
points and hence the division in Fig. 1 could give an idea about the behavior
of $M=M(x)$ as well.

\section {Analytical Expression for the Shock Location and Behavior in Parameter Space}

A black hole accretion flow being transonic, it
must satisfy two sonic point conditions at the
cost of one extra unknown, namely, the the sonic point.
Because of this extra condition, out of the three
constants of motion, namely, ${\cal E}$, ${\dot M}$
and $\la$, only two are to be supplied as free parameters.
While computing shock locations, we used past experience
derived from numerical methods, that only one of the shocks,
namely $x_{s3}$ (C89 notation) is stable, and accordingly
our procedure as delineated below attempts to compute
only this location.

The flow will have a shock only at the point where the
shock invariant condition is satisfied. Simplifying the
shock invariant relation (Eq. 18), we obtain,
\be
2(\ga-1)(M^{2}_{+}+M^{2}_{-})-[(3\ga-1)^{2}-2(3\ga-1)(\ga-1)]M^{2}_{+}M^{2}_{-}+4=0.
\ee
We consider a relativistic flow with $\ga=4/3$ so that the polytropic index $n=3$.
Then from Eq. (22), we obtain,
\be
2(M^{2}_{+}+M^{2}_{-}) -21M^{2}_{+}M^{2}_{-} +12=0.
\ee

We now expand the post-shock Mach number $M^{2}_{+}$ by a
polynomial which must satisfy the following conditions:

(a) $\frac {dM}{dx}$ is zero at the central `O'-type sonic point. This is a
general property of the flow (vide discussion at the end of last Section).

(b) Mach no. ($M_+$) at the location of the
middle sonic point must match with that derived
from approximate analytical solution obtained using the energy equation (Eq. 2 and 4a).

(c) Solution must pass through the position where $\frac {dM}{dx}$ is $\infty$
and a good guess of this the location (say, from the location of the sonic
points) is known.

Similarly, we expand $M^{2}_{-}$ by a polynomial which must satisfy
the following conditions:

(a) $\frac {dM}{dx}$ is zero at the outer sonic point (location of which is
already determined above). This is a general condition (vide discussion
at the end of the previous Section).

(b) Mach No. at the outer sonic point must match with the analytical value
obtained from the Sonic point condition (Eq. 12).

(c) Mach No. ($M_-$) at the location of the middle sonic point 
must match with the analytical value derived from approximate 
analytical solution obtained using the energy equation (Eq. 2 and 4a).

Keeping in mind that an algebraic equation which is beyond quartic
cannot be solved analytically [see, \citet{as70}], we expand $M^2_\pm$ as quadratic
equations so that Eq. (23) may become quartic.
We shall show a posteriori that such an assumption introduces a very small
and tolerable error in our computation.

If $x_s$ denotes the shock location, we assume,
\be
M^{2}_{\pm}=\sum^{2}_{q=0} A_{[q,\pm]}x^{q}_{s}
\ee
where, $A_{[q,\pm]}$ are constant co-efficients to be determined 
from the conditions mentioned above. We find them to be,

$A_{[2,+]}=\frac {1-\left(M^2_{+}\right)_{mid}}
{(x_{inf}-x_{mid})^2}$,

$A_{[1,+]}=-2x_{mid}A_{[2,+]}$,

$A_{[0,+]}=1+\left( 2x_{mid}x_{inf}-x^2_{inf}\right)A_{[2,+]}$,

$A_{[2,-]}=\frac {\left( M^2_{-}\right)_{mid}-1}{(x_{out}-x_{mid})^2}$,

$A_{[1,-]}=-2x_{mid}A_{[2,-]}$,

$A_{[0,-]}=1+x^2_{out}A_{[2,-]}$,

\noindent where, $x_{mid}$, $x_{inf}$ and $x_{out}$ are the
middle (O-type) sonic point, the position where first derivative
of Mach No. is infinity and the outer sonic point respectively.

We now substitute the above expression (Eq. 24) for the Mach number
in the Mach invariant relation (Eq. 23) to obtain the following
algebraic equation:

\begin{eqnarray}
{\cal A}x^{4}_{s}+{\cal B}x^{3}_{s}
+{\cal C}x^{2}_{s}
+{\cal D}x_{s}+{\cal F}=0,
\end{eqnarray}

\noindent where,

${\cal Y}=\left(3\ga -1\right)\left(\ga+1\right)$, 

${\cal Z}=2(\ga-1)$,

${\cal A}={\cal Y}A_{[2,+]}A_{[2,-]}$,

${\cal B}={\cal Y}\left(A_{[1,+]}A_{[2,-]}+A_{[2,+]}
A_{[1,-]}\right)$,

${\cal C}={\cal Y}\left(A_{[0,+]}A_{[2,-]}+A_{[1,+]}A_{[1,-]}+
A_{[2,+]}A_{[0,-]}\right)
-{\cal Z}\left(A_{[2,+]}+A_{[2,-]}\right)$,

${\cal D}={\cal Y}\left(A_{[0,+]}A_{[1,-]}+A_{[1,+]}
A_{[0,-]}\right) -{\cal Z}\left(A_{[1,+]}+A_{[1,-]}\right)$,

${\cal F}={\cal Y} A_{[0,+]}A_{[0,-]}-{\cal Z}\left(A_{[0,+]}+
A_{[0,-]}\right)-4$.

We solve this equation for $X_s$ analytically using the same procedure
as in \S4 (details in Appendix A). We denote the discriminant $D$ by $D_s$,
and $Q$ and $R$ values as $Q_s$ and $R_s$ respectively
for our discussion on the parameter space behavior of shocks.

In Figure 2, we redraw the parameter space as in Figure 1, but consider
the formation of shocks alone.  We find that $Q_s>0$ produces no shock
from above, and $Q_s=0$ with $R_s\ne 0$ gives the boundary of the weakest shock
(shocks with unit compression ratio). This boundary, though
obtained using our approximate analytical method, generally coincides 
with the dotted curve of Figure 1. The edge of the boundary
is obtained with an extra condition $R_s=0$. Thus, $R_s$
progressively decreases towards the edge along the dashed curve.
This edge (denoted by $D_s=R_s=0$) ought to have coincided with
the cusp of the $D_c=0$ curve drawn for the sonic point
(see, Fig. 1 also), had the analytical method been exact. A small
shift is the evidence that a small error is present at this corner of the parameter space.
We also provide the region of the oscillating shocks ($Q_s<0$ and $D_s<0$). Here, the
shock location is imaginary and therefore shock continuously
oscillates back and forth causing a very interesting astrophysical
effect known as Quasi-Periodic Oscillations (QPOs).
This would be discussed in the next Section. The boundary
between the shock and no-shock region from below is
denoted by the dashed curve marked by $D_s=0$. Below the
no-shock region where the energy and angular momentum
of the flow are very low, the flow has only one sonic point.

In Figure 3, we draw shock location  $x_s$ (along Y-axis)
as a function of energy (along X-axis). Different curves
are drawn for different specific angular momentum
of the flow. The rightmost one is for $\lambda=1.51$
and the leftmost one is for $\lambda=1.84$, interval of $\la$ is
$0.01$. As angular momentum is increased, the shock is located
farther from the black hole. Comparing the locations from those
obtained analytically, one notices that the same location is obtained
for a specific angular momentum slightly more
($\sim 3$\%) than that used in the numerical method.
We therefore believe that the results obtained are
very much reliable.

In Figure 4, we present a comparison of the boundary of
the parameter space for which standing shocks may form
as obtained by our analytical solution (shaded region) and
by the numerical means (solid curve) existing in the
literature (C89). The agreement is very good except in
a region near the cusp (as also noted while discussing Fig. 2).
Since very little parameter space is involved at this edge,
we think that this small mismatch is tolerable.

\section{Astrophysical Applications}

Even though a black hole has no hard surface, it is remarkable
that matter forms standing shocks around it in the same way as a
shock is formed when a supersonic flow encounters a hard boundary.
Shock waves heat up a gas and puff it up. This post-shock region
intercepts soft photons from
the pre-shock matter, particularly from the Keplerian disk  which
is located on the equatorial plane (CT95).
In this scenario, the nature of the Comptonized radiation depends on the amount
of matter in the sub-Keplerian and in the Keplerian flow: if the intensity
of soft photon is very low, they cannot cool down the post-shock
region by inverse Compton process and the spectrum remains very
hot. On the other hand, if the intensity of soft photon is
very high (i.e., the Keplerian rate is large), they cool down
the post-shock region to the extent that the shock could not be
sustained (pressure balance condition breaks down). This
produces a soft state spectrum with a hard tail due to bulk motion
Comptonization (CT95). There are several models
in the literature which perhaps explain the soft/hard states. However,
no model other than CT95 explains the power-law hard-tail in the soft state. 
Similarly, when the question of Quasi-Periodic Oscillations (QPOs)
come, shock-oscillation model turns out to be  a sufficiently satisfactory one
(CM00).

When the parameters fall in the `no shock' region of Fig. 2,
shock location becomes imaginary. However, three sonic points
are still present and the entropy of the flow at the inner sonic point
continues to be higher compared to that at the outer sonic point.
In this case, shock starts oscillating with a time period ($T_s$) comparable
to the infall time from the post-shock region \citep{rcm97}.
Even when shocks form, if the infall time-scale turns out to be
comparable to the cooling time, then the resonance condition is satisfied
\citep{msc96,cm00}
and shocks oscillate in time scales of:
$$
T_s\sim x_s/v_s ,
$$
where, $v_s \sim R^{-1}x_s^{-1/2}$ is the infall velocity
and $R$ is the compression ratio at the shock (easily
obtained analytically from our equations). Observed QPO
frequencies are comparable to $1/T_s$.

When a mixture of Keplerian and sub-Keplerian matter is accreted, it is
easy to obtain the parameters ${\lambda}$ and ${\cal E}$ in terms of
the Keplerian (${\dot M}_{d}$) and  sub-Keplerian (${\dot M}_h$)
accretion rates. Suppose, the viscosity parameter
is such that the flow is deviated from a Keplerian disk
at $x=x_K$ where its energy and angular momentum were ${\cal E}_d$
and $\lambda_K$ respectively. Further suppose that
the sub-Keplerian halo has a constant energy ${\cal E}\sim 0\sim {\cal E}_h$
and constant angular momentum ${\lambda}_h$, then the average
angular momentum and energy of the transonic flow would be,
$$
<{\cal E}> = \frac{{\dot M}_d  {\cal E}_d + {\dot M}_h {\cal E}_h}{{\dot M}_d +{\dot M}_h}
$$
$$
<\lambda> = \frac{{\dot M}_d  \lambda_d + {\dot M}_h \lambda_h}{{\dot M}_d +{\dot M}_h}
$$
It is easy to compute the shock location of the resultant flow using our formalism
given above. 

It is to be noted from Fig. 3 that shock solutions are allowed only if the
specific energy is positive. In other words, if a flow deviates from a
cool Keplerian disk on the equatorial plane, the flow cannot have shocks
as the specific energy in such a flow would be negative unless
this flow is mixed with a substantial amount of sub-Keplerian matter with a
positive energy.  Typically, ${\dot M}_h >> {\dot M}_d$ and even with a small energy, the specific 
energy of the mixture becomes positive, giving rise to shocks and (unbound) winds.
In the case when magnetic dissipation is present, flow
energy could increase to a positive value and a solution with a shock would be allowed.
The prospect of magnetic energy dissipation has been discussed by several
workers in the literature \citep{s73,bb76,b98}. Briefly, since the magnetic field rises as
$B_r \propto r^{-2}$ and therefore magnetic pressure rises as
$P_{mag} \propto r^{-4}$ while the gas pressure in the sub-Keplerian matter
goes as $P_{gas} \propto r^{-5/2}$, magnetic field in excess of the equipartition
value would escape from the disk buoyantly and may dissipate at the atmosphere
as in the case of the Sun. If the flow has specific energy ${\cal E}_h$
at, say $r=100r_g$ where the flow was in equipartition,
then at the shock, the energy would be at least $2^4 =16$ times
larger if {\it all} the magnetic energy is dissipated into the flow.
Thus, a basically free-fall matter of ${\cal E} \sim 10^{-4}$
would have an energy $\sim 10^{-3}$ and thus a shock at a few tens of
Schwarzschild radii would be expected. 

An accreting flow can intercept hard X-rays emitted at the inner edge
and this pre-heating effect need not be negligible. For instance, 
a flow emitting isotropically with $6\%$ efficiency will definitely
intercept $\theta \sim \Theta_{in}/4\pi$ fraction of radiation
in between the shocked region and the Keplerian disk. Assuming $\theta \sim 0.1$,
the energy deposition due to pre-heating is $0.006$ which is significant.
This would energize Keplerian matter as well and shocks in the sub-Keplerian
flow would be expected.

\section{Concluding Remarks}

So far in the astrophysical literature, existence of 
shocks in accretion flows has been indicated by steady and time-dependent
numerical simulations. Study of these 
standing and oscillatory shocks in an accretion flows
has been shown to be of great importance. Presently we show that the shocks
could be studied completely analytically, at least in the case of
thin, axisymmetric, inviscid flows with positive energy. We note that shock locations
vary with flow parameters in a simple way --- they form farther from a black hole
when angular momentum is increased. This proves that they are mainly centrifugal
pressure supported.

Given that the shocks, especially the standing shocks are the ideal locations
at which a flow is heated up, hard X-rays are produced from the post-shock
region after the soft-photons are processed by the flow by
inverse-Comptonization process. Thus, the spectral states and the time-dependent
behavior of the hard X-rays are directly related to the behavior of this
region. For instance, CT95 computes steady state spectra with post-shock 
region as the source of hot, Comptonizing electrons. CM00 establishes that
QPOs are due to oscillations of this region, since only hard X-rays are seen to
exhibit QPOs. We therefore believe that shocks should be an important
ingredient of an accreting system. However, if the disk is cool and Keplerian
very far away, the specific energy must be negative. So the problem is not whether
shocks should exist, the problem seems to be how to energize Keplerian matter
as it becomes sub-Keplerian, by, say, magnetic energy 
dissipation, pre-heating etc. These work is in progress and would be reported elsewhere.

From the observers point of view, our work could also be useful, since the steady spectra,
QPO frequencies, etc. are, in principle, determined analytically
from a few free physical parameters. In future, we shall apply our 
understanding to obtain spectral properties and QPO behaviors more quantitatively.

\acknowledgments

This work is partly supported by a project (Grant No. SP/S2/K-14/98)
funded by Department of Science and Technology (DST).

{}

\appendix

\section{Method of the Analytical Solution}

The procedure of obtaining an analytical solution of a
quartic equation,
$$
q^{4}+b_{1}q^{3}+b_{2}q^{2}+b_{3}q+b_{4}=0
\eqno{A.1}
$$
is to first  obtain a solution of the following cubic equation:

\noindent
$$
p^{3}+a_{1}p^{2}+a_{2}p+a_{3}=0,
\eqno{A.2}
$$
where,\\
$a_1=-b_2$, $a_2= b_{1}b_{3}-4b_{4}$, and $a_3=4b_{2}b_{4}-
b^{2}_{3}-b^{2}_{1}b_{4}$ .

\noindent Let,

$Q=\frac {3a_{2}-a^{2}_{1}}{9}$, $R=\frac {9a_{1}a_{2}
-27a_{3}-2a^{3}_{1}}{54}$,\\

$S=\sqrt[3]{R+\sqrt{Q^{3}+R^{2}}}$, $T=\sqrt[3]
{R-\sqrt{Q^{3}+R^{2}}}$.

\noindent The {\it discriminant} is defined as

$$
D=Q^{3}+R^{2}.
\eqno{A.3}
$$

\noindent If $D>0$, one root is real and two roots are
complex conjugate. In this case, the real solution is,
$$
p_{1}=S+T-\frac {1}{3}a_{1}.
\eqno{A.4}
$$

\noindent If $D=0$, all roots are real and at least two are equal.

\noindent If $D<0$, all roots are real and unequal. They are:
$$
p_{1}=2\sqrt{-Q}\cos(\frac {1}{3}\theta)-\frac {1}{3}a_{1},
\eqno{A.5}
$$

$$
p_{2}=2\sqrt{-Q}\cos(\frac {1}{3}\theta+120^{o})-\frac {1}{3}a_{1},
\eqno{A.6}
$$
and,
$$
p_{3}=2\sqrt{-Q}\cos(\frac {1}{3}\theta+240^{o})-\frac {1}{3}a_{1},
\eqno{A.7}
$$
where, $\cos \theta=R/\sqrt{-Q}$.

One can now write  a quadratic equation using
any one of the real solutions of the cubic equation
(see, Spiegel, 1968) as follows:
$$
z^{2}+\frac {1}{2} \{ b_{1} \pm \sqrt {b^{2}_{1}-4b_{2}+4p_{1}} \}z+
\frac {1}{2} \{ p_{1} \mp \sqrt {p^{2}_{1}-4b_{4}} \}=0.
\eqno{A.8}
$$
This is a quadratic equation which can be solved easily. Since we applied this procedure
both of the sonic points and shocks, we denoted quantities like, $D$,$Q$ and $R$
by $D_c$,$Q_c$ and $R_c$ for sonic (critical) points and $D_s$,$Q_s$ and $R_s$ for
shocks respectively.
\newpage

\figcaption[fig1.eps]{Division of the parameter space as spanned by the pair
${\cal E}, \la$ according to the number of sonic points.
Solid curve represents $D_c=0$ which divides the
region into $D_c>0$ (one sonic point) and $D_c<0$ (three sonic points)
regions. The plot is for $\gamma=4/3$. For $\gamma \gsim 1.5$, $D_c>0$ always,
suggesting that no shocks are possible in a flow with vertical equilibrium.
The dashed curve further divides the region into two regions where
entropy accretion rate ${\dot {\cal M}}$ at the two saddle type sonic points
behave differently (inner point is denoted by $i$ and outer point is
denoted by $o$). }

\figcaption[fig2.eps]{Division of the parameter space as spanned by the pair
${\cal E}, \la$ according to whether shocks would form or not.
Solid curve represents $D_c=0$ as in Fig. 1. Dashed curve
($D_s=0$) surrounds the region with shocks in accretion. When $D_s<0$ and
yet, there are three sonic points, shocks are oscillatory, giving rise to
quasi-periodically varying hard X-rays.}

\figcaption[fig3.eps]{
Variation of shock location ($x_s$ along y-axis) with specific energy
(${\cal E}$ along x-axis) of the flow. Each curve is drawn for a
specific angular momentum $\lambda$. From right to left curved are
drawn for $\lambda=1.51,~ 1.52,~1.53, ...$ till $1.84$ respectively.
For a given specific energy ${\cal E}$, shock location increases
with increasing centrifugal force (through $\lambda$). Similarly,
for a given $\lambda$, shock location increases with energy.
}

\figcaption[fig4.eps]{
Comparison of the boundary of the parameter space  in the
(${\cal E}, \lambda$) plane using the numerical and the
analytical methods. Except for the region near the cusp (upper left corner)
the agreement is very strong. }

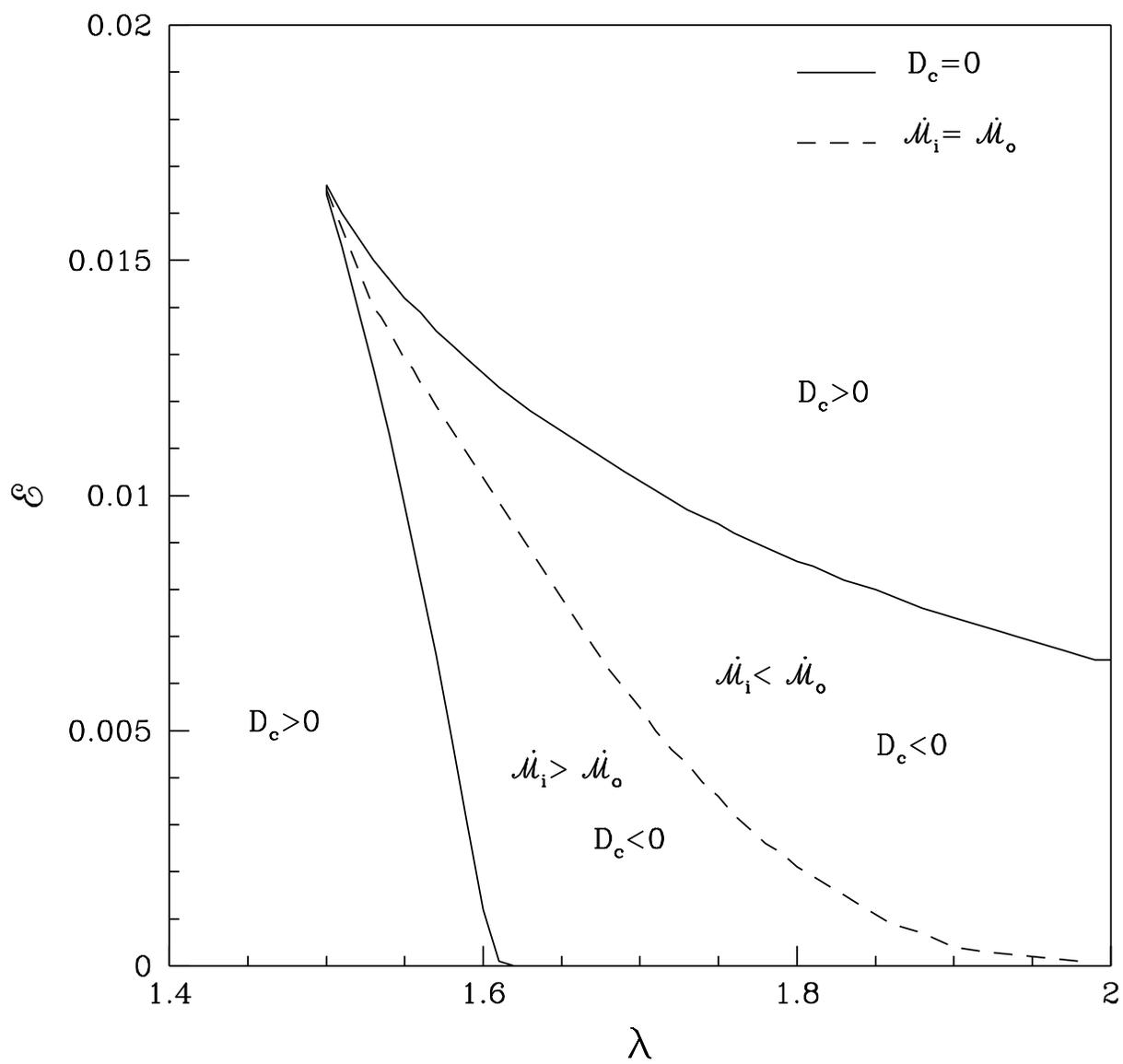
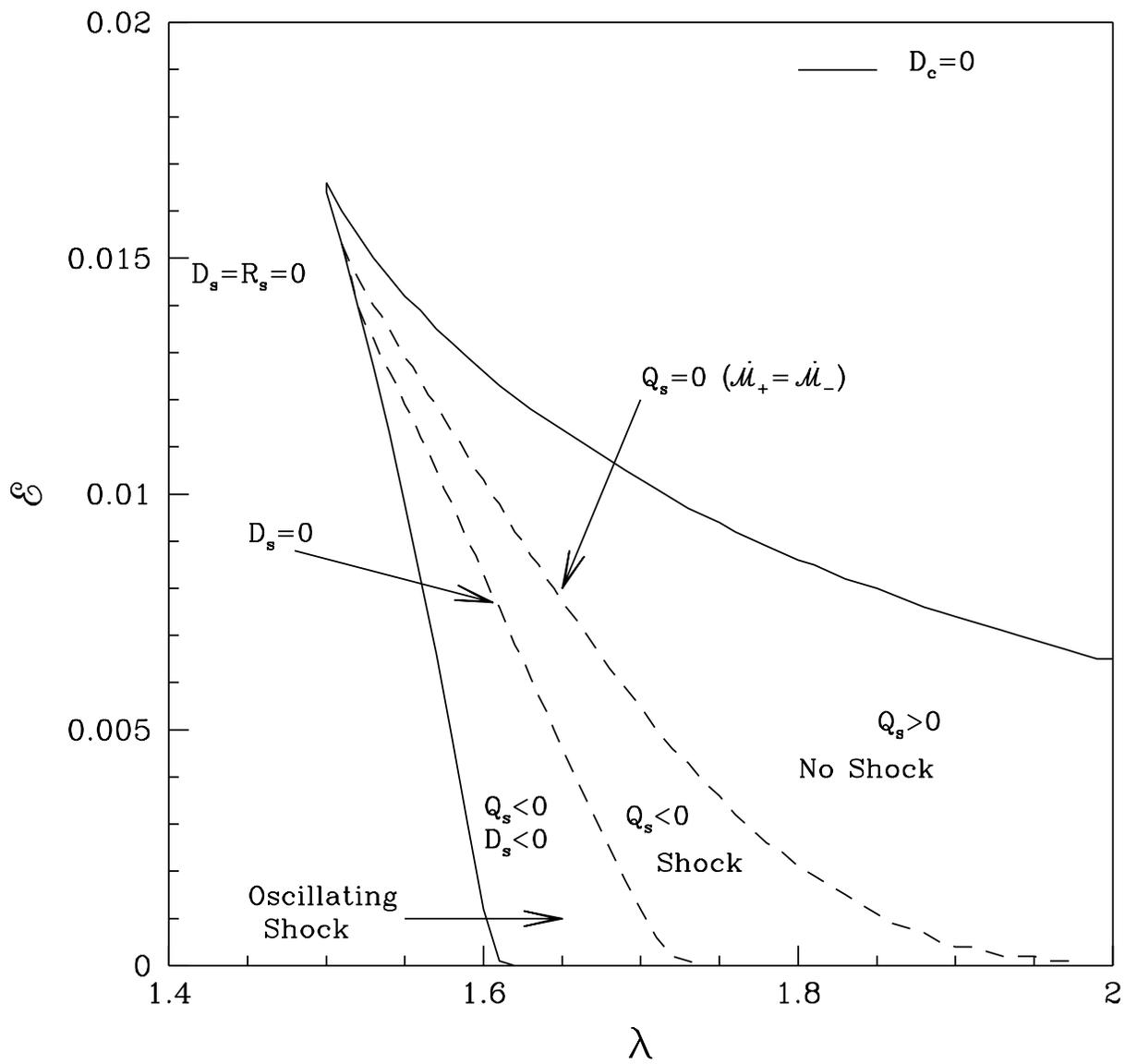
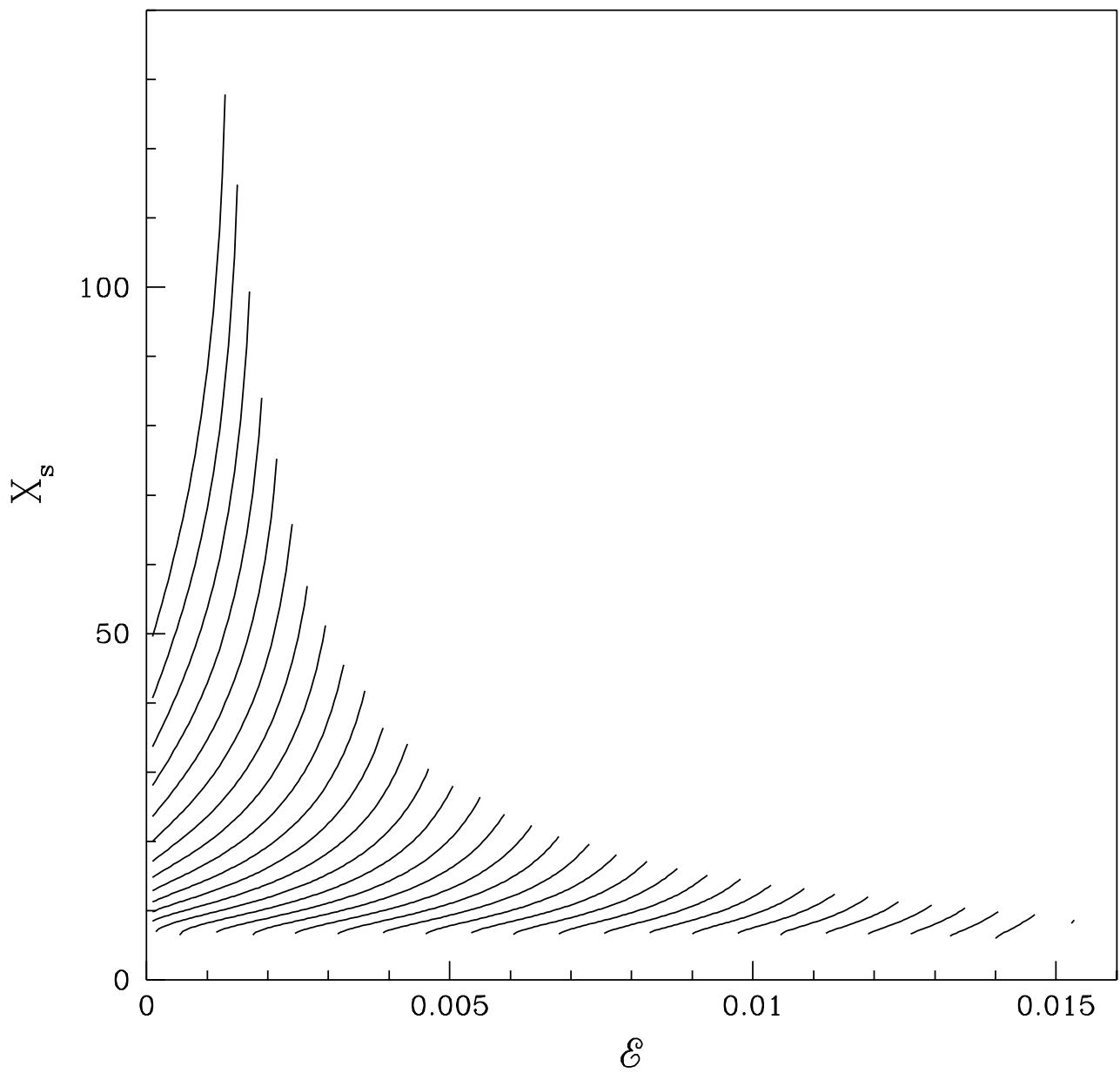
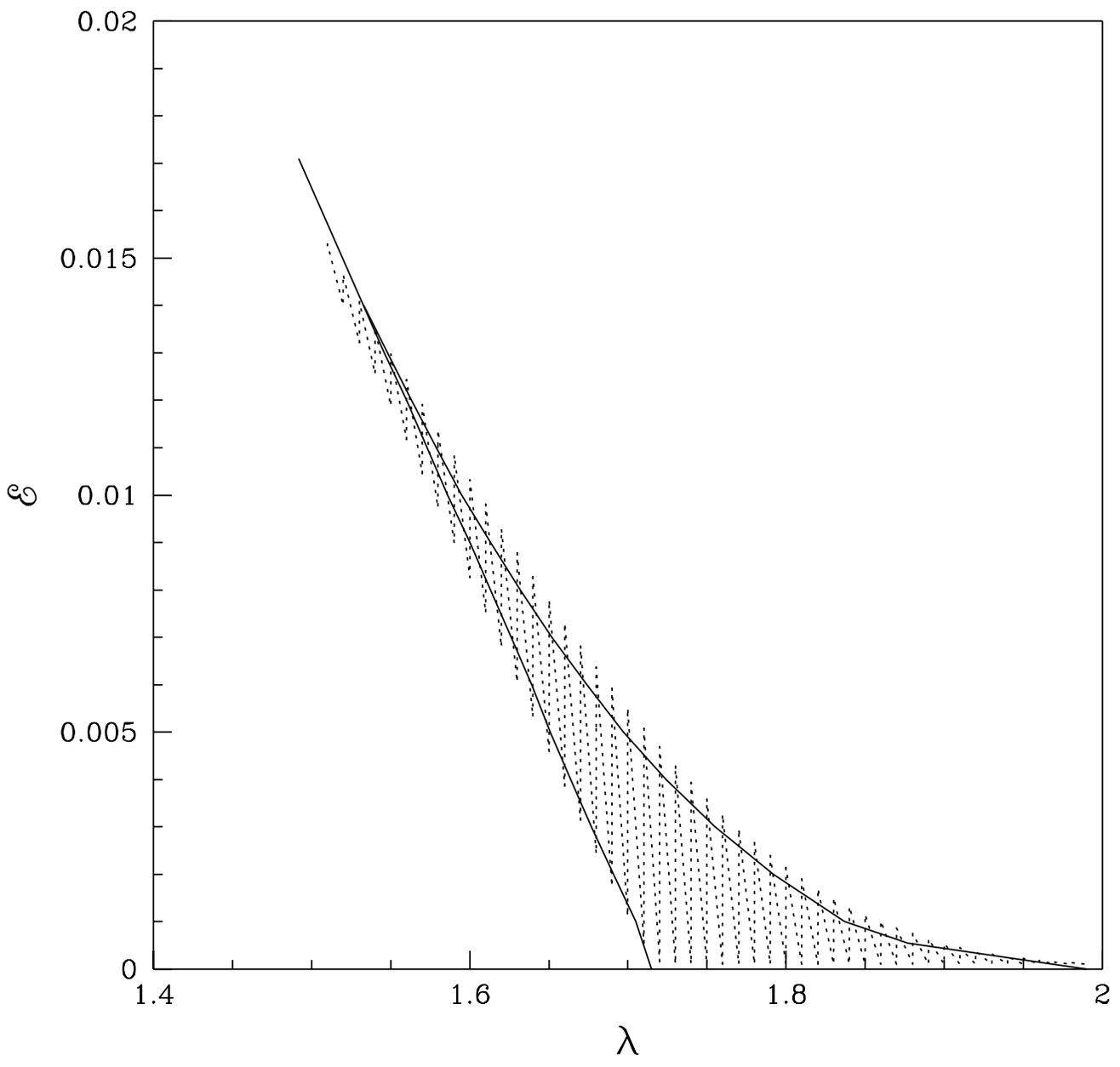
\end{document}